\documentclass[aps,prb,preprint,superscriptaddress]{revtex4-1}

\usepackage[utf8]{inputenc}
\DeclareUnicodeCharacter{00B0}{$\,^\mathrm{o}$}
\usepackage{lmodern}
\usepackage{graphicx}
\usepackage{amsmath}
\usepackage{hyperref}
\usepackage{color}
\usepackage[normalem]{ulem}
\begin{document}

%Title of paper
\title{Charge density wave and superconductivity competition in Lu$_5$Ir$_4$Si$_{10}$ : a proton irradiation study}

\author{Maxime Leroux}
\altaffiliation{Present address: Laboratoire National des Champs Magnétiques Intenses (CNRS, EMFL, INSA, UGA, UPS), Toulouse 31400, France}
\affiliation{Materials Science Division, Argonne National Laboratory, Argonne, USA}

\author{Vivek Mishra}
\altaffiliation{Present address: Kavli Institute for Theoretical Sciences, University of Chinese Academy of Sciences, Beijing 100190, China}
\affiliation{Materials Science Division, Argonne National Laboratory, Argonne, USA}

\author{Christine Opagiste}
\affiliation{Institut NÉEL, CNRS, Univ. Grenoble Alpes, Grenoble 38000, France}

\author{Pierre Rodière}
\affiliation{Institut NÉEL, CNRS, Univ. Grenoble Alpes, Grenoble 38000, France}

\author{Asghar Kayani}
\affiliation{Department of Physics, Western Michigan University, Kalamazoo, USA}

\author{Wai-Kwong Kwok}
\affiliation{Materials Science Division, Argonne National Laboratory, Argonne, USA}

\author{Ulrich Welp}
\affiliation{Materials Science Division, Argonne National Laboratory, Argonne, USA}

\date{\today}

\newcommand{\Tc}{$T_\mathrm{c}$}
\newcommand{\TCDW}{$T_\mathrm{CDW}$}

\newcommand{\LBCOx}{La$_{2-x}$Ba$_x$CuO$_4$}
\newcommand{\LNSCOx}{La$_{1.6-x}$Nd$_{0.4}$Sr$_x$CuO$_4$}
\newcommand{\LBCO}{La$_{1.875}$Ba$_{0.125}$CuO$_4$}
\newcommand{\LBSCOx}{La$_{1.875}$Ba$_{0.125-x}$Sr$_x$CuO$_4$}
\newcommand{\YBCOd}{YBa$_2$Cu$_3$O$_{7-\delta}$}
\newcommand{\BSCCOx}{Bi$_2$Sr$_2$CaCu$_2$O$_{8+x}$}
\newcommand{\HBCOd}{HgBa$_2$CuO$_{4+\delta}$}
\newcommand{\NCCOx}{Nd$_{2-x}$Ce$_x$CuO$_4$}
\newcommand{\LIS}{Lu$_5$Ir$_4$Si$_{10}$}

\begin{abstract}
Real-space modulated Charge Density Waves (CDW) are an ubiquituous feature in many families of superconductors.
In particular, how CDW relates to superconductivity is an active and open question that has recently gathered much interest since CDWs have been discovered in many cuprates superconductors. 
Here we show that disorder induced by proton irradiation is a full-fledged tuning parameter that can bring essential information to answer this question as it affects CDW and superconductivity with different and unequivocal mechanisms.
Specifically, in the model CDW superconductor \LIS\ that develops a 1D CDW below 77\,K and s-wave superconductivity below 4\,K, we show that disorder enhances the superconducting critical temperature \Tc\ and $H_\mathrm{c2}$ while it suppresses the CDW.
Discussing how disorder affects both superconductivity and the CDW, we make a compelling case that superconductivity and CDW are competing for electronic density of states at the Fermi level in \LIS, and we reconcile the results obtained via the more common tuning parameters of pressure and doping.
Owing to its prototypical, 1D, Peierls type CDW and the s-wave, weak-coupling nature of its superconductivity, this irradiation study of \LIS\ provides the basis to understand and extend such studies to the more complex cases of density waves and superconductivity coexistence in heavy fermions, Fe-based or cuprates superconductors. 

\end{abstract}

% insert suggested PACS numbers in braces on next line
% insert suggested keywords - APS authors don't need to do this
%\keywords{}

%\maketitle must follow title, authors, abstract, \pacs, and \keywords
\maketitle

\section{Introduction}

A charge density wave (CDW) is a spatial modulation of the electronic density of states which opens a gap at the Fermi level. CDW can arise from electronic instabilities such as Fermi surface nesting in low-dimension metals\cite{gruner2000density,MonceauCDWreview2012} or a peak in electron-phonon coupling\cite{ZhuCDWclassification}. This charge modulation is usually accompanied by a periodic lattice distortion, via the electron-phonon coupling. An analog modulation known as a spin density wave (SDW) also exists for the electronic spin density\cite{gruner2000density,MonceauCDWreview2012}.
The presence of real-space modulated CDW or SDW is a feature of many families of superconductors\cite{MonceauCDWreview2012,Gabovich_2001Review}.

Recently, CDWs have been found to be ubiquitous in many cuprates superconductors, whether in hole-doped \LBCOx\cite{Tranquada1995stripenature}, \YBCOd\cite{WuJulien2011CDWNMR,Wu2013,Wu2015,LeTacon2013IXSYBCONatPhys,Chang2012YBCOcompetingCDW,Gerber3DCDWYBCOhighB,JangYBCOCDWhighfield}, \BSCCOx\cite{NetoGuTacon2014Science}, \HBCOd\cite{Tabis2014NatComHgBCO_cdw} or in electron doped \NCCOx\cite{daSilvaNetoCDWelecdopedCuprate}.
% this has renewed the interest in investigating superconductivity in the presence of density waves\cite{FradkinTranquada2015RoMP}.
Other examples include: Fe-based superconductors in which superconductivity seems to compete with spin density waves\cite{Fernandes2012Tcupwithdisorder,Mishra2015SDW}; heavy fermion compounds where the SDW appears linked to d-wave superconductivity\cite{Duk2016_PRX_CeCoIn5}; transition metal dichalcogenides where the CDW is well-known to compete with superconductivity in 2H-TaS$_2$ and 2H-TaSe$_2$\cite{Mutka1983PRB_CDW_SC_irrad} but 1T-TiSe$_2$ has been proposed as an excitonic superconductor enhanced by the CDW\cite{KusmartsevaPRL2009TiSe2CDW_SC}; finally organic superconductors also exhibit superconductivity in coexistence with density waves (e.g. (TMTSF)$_2$PF$_6$)\cite{JeromeACSreview2004OrganicSC}. 
Thus, whether CDW competes with\cite{LeTacon2013IXSYBCONatPhys,Chang2012YBCOcompetingCDW,LerouxPNAS2019} or on the contrary are a key ingredient in explaining the origin of cuprates' high temperature superconductivity\cite{BergPRL2007LBCOTcT3DTKT,BergTranquada2009NJoP}, the relation between density waves and superconductivity is an active and open question.\cite{FradkinTranquada2015RoMP}

Among superconductors with CDW,
Lu$_5$Ir$_4$Si$_{10}$ is a well established case of s-wave superconductivity coexisting with a standard Peierls-type CDW\cite{Shelton1986LIS_P_rho_Cp,Yang91,JungMiglioriPressure2003,Mansart2012LIS_femtosec_and_DFT}.
This compound possesses a first order CDW transition below $T_\mathrm{CDW}=77$\,K\cite{BettsMigliori2002LIS_RUS,SAINTPAULJPCS2020} and it also becomes superconducting below $T_\mathrm{c}=4$\,K\cite{Leroux2013LIS_SCprop}. The CDW develops on 1D Lutetium atom chains along the c-axis, following the nesting mechanism\cite{Mansart2012LIS_femtosec_and_DFT}, with clear signatures in electrical transport\cite{Shelton1986LIS_P_rho_Cp,JungMiglioriPressure2003}, x-ray\cite{Becker1999}, specific heat\cite{Shelton1986LIS_P_rho_Cp,Ramakrishnan_2017,Becker1999} or elastic constants\cite{BettsMigliori2002LIS_RUS,SAINTPAULJPCS2020}.
The CDW is commensurate with a periodicity of seven unit cells as evidenced by x-ray diffraction\cite{Becker1999}
The CDW gaps an estimated 36\% of the density of states at the Fermi level as determined from resistivity and specific heat measurements\cite{Shelton1986LIS_P_rho_Cp}, with more recent optical estimates ranging from 16\%\cite{Tediosi2009prbopticLIS} to 30\%\cite{Mansart2012LIS_femtosec_and_DFT}. 
The effect of pressure points to a competition scenario: from 0 to 2\,GPa \TCDW\ decreases continuously and \Tc\ is constant, but above 2\,GPa the CDW suddenly vanishes and \Tc\ jumps from 4 to 9\,K\cite{Shelton1986LIS_P_rho_Cp}.
Chemical doping also points to a competition scenario: the CDW state is suppressed and \Tc\ continuously increases up to at least 6\,K for increasing doping\cite{Yang91,SinghPRB2005,Leroux2013LIS_SCprop}. We note that the increase of \Tc\ is binary in the former case, but progressive in the latter. To explain this differing behavior of \Tc\ between pressure and doping, it has been proposed that \LIS\ presents a sharp feature in the electronic density of states just above the Fermi level.\cite{Yang1986sharpfeature}

In this article, we establish disorder induced by irradiation as a full-fledged axis in the phase diagrams of superconductors,
via an extensive study of this model compound \LIS. In particular, we evidence the mechanism through which proton irradiation acts as a tuning parameter suppressing the CDW in favor of superconductivity and we show how this tuning parameter brings its own set of unique information on superconductivity-CDW competition.
In sharp contrast with the expected effect of disorder on superconductivity, we observe an \emph{increase} of \Tc\ after irradiation.
%%%NEW
Proton irradiation produces cascade-type clusters of defects, which are typically a few nanometer in diameter, along with a small fraction of point defects.
%%%
This irradiation induced disorder strongly suppresses the CDW and broadens its transition, thus revealing the precise mechanism of CDW suppression. Moreover the increase of $H_{c,2}$ with disorder reveals that the channel for competition between CDW and superconductivity is the electronic density of states at the Fermi level.
These results make a compelling case that reconciles how CDW and superconductivity competes in \LIS\ with pressure, doping and disorder.
This extensive set of results in a BCS s-wave compound with a prototypical 1D CDW of the Peierls type provides the basis to pursue such irradiation studies in the more complex cases of density wave coexistence in heavy fermions, Fe-based or cuprates superconductors\cite{LerouxPNAS2019}.

The article is organized as follows:
Part.~\ref{MatMeth} presents the materials and methods.
Part.~\ref{disorderTc} presents how irradiation induced disorder raises \Tc\ and reduces \TCDW.
Part.~\ref{magfield} presents the evolution of $H_\mathrm{c,2}$ as a function of disorder.
Part.~\ref{Tcpresdopdis} reconciles the different evolutions of \Tc\ with pressure, doping and disorder by discussing how they relate to the CDW suppression mechanisms.

%%%%%%%%%%%%%%%%%%%%%%%%%%%%%%%%%%%%%%%%%%%%%%%%%%%%%
%%%%%%%%%%%%%%%%%%%%%%%%%%%%%%%%%%%%%%%%%%%%%%%%%%%%%
%%%%%%%%%%%%%%%%%%%%%%%%%%%%%%%%%%%%%%%%%%%%%%%%%%%%%
% part II
\section{Materials and methods}\label{MatMeth}
Lu$_5$Ir$_4$Si$_{10}$ has a tetragonal unit cell with lattice parameters $a = 12.484(1)$ and $c=4.190(2)$\,\AA\cite{OpagisteLeroux2010LISwhiskers} and space group symmetry P4/mbm. 
The CDW forms along the c-axis on quasi 1D chains of lutetium atoms.
The samples are high quality single crystals that grow in needle shape along the c-axis, and have been characterized previously\cite{OpagisteLeroux2010LISwhiskers,Leroux2013LIS_SCprop}. 
We used the tandem van de Graaf accelerator at Western Michigan University to irradiate a sample several times with 4\,MeV protons. This sample has dimensions $10 \,\mathrm{\mu m} \times 65 \,\mathrm{\mu m} \times 500 \,\mathrm{\mu m}$ $\mathrm{(a\times b \times c)}$. The 10\,$\mu$m thickness of the sample ensures uniform irradiation damage and negligible proton implantation, as SRIM calculations\cite{Ziegler2010SRIM} show the projected range of protons is 67\,$\mu$m in these conditions.

For the irradiations, the sample is mounted onto an aluminum sample holder that allows for linear and rotational motion.  In order to avoid heat damage of the sample we use a relatively low beam current of 500 nA and a cooling stage that maintains the sample at -10$^\circ$C during irradiation. The incident proton beam of 4.7 mm diameter is homogenized by passing through a 1 $\mu$m gold foil placed at 240 mm upstream from the sample. The beam is defined through a 7.8 mm aperture placed at 40 mm upstream. This set-up is calibrated with the help of a Faraday cup placed down-stream from the sample which captures all protons passing the aperture while the sample is moved out of the beam path. The sample is electrically connected to the sample holder and the irradiation chamber which, in turn, is isolated from all other electronics and from the beam pipe through plastic rings.  This approach allows to accurately determine the irradiation dose by integrating the current from the chamber, not affected by spurious effects due to the emission of secondary electrons. The sample was irradiated in four sessions at Western Michigan University to a rather high cumulative dose of 12x10$^{16}$ p/cm$^2$ (protons per cm$^2$). Such a high dose is known to start to degrade some superconducting properties in several families of superconductors\cite{LerouxPNAS2019}.
After each irradiation we measured the resistivity using a Keithley 2182 voltmeter and 6221 current source, with currents ranging from 50\,$\mu$A to 1\,mA, in an helium 4 cryostat with a 7\,T magnet. Contacts were made with sputtered platinum and silver epoxy Epotek H20E. Typical contact values are $\lesssim 5\,\Omega$. The voltage contacts were spaced 200\,$\mu$m apart along the c-axis. 

%%%%%%%%%%%%%%%%%%%%%%%%%%%%%%%%%%%%%%%%%%%%%%%%%%%%%
%%%%%%%%%%%%%%%%%%%%%%%%%%%%%%%%%%%%%%%%%%%%%%%%%%%%%
%%%%%%%%%%%%%%%%%%%%%%%%%%%%%%%%%%%%%%%%%%%%%%%%%%%%%
% part III
\begin{figure}
\includegraphics[width=0.5\textwidth]{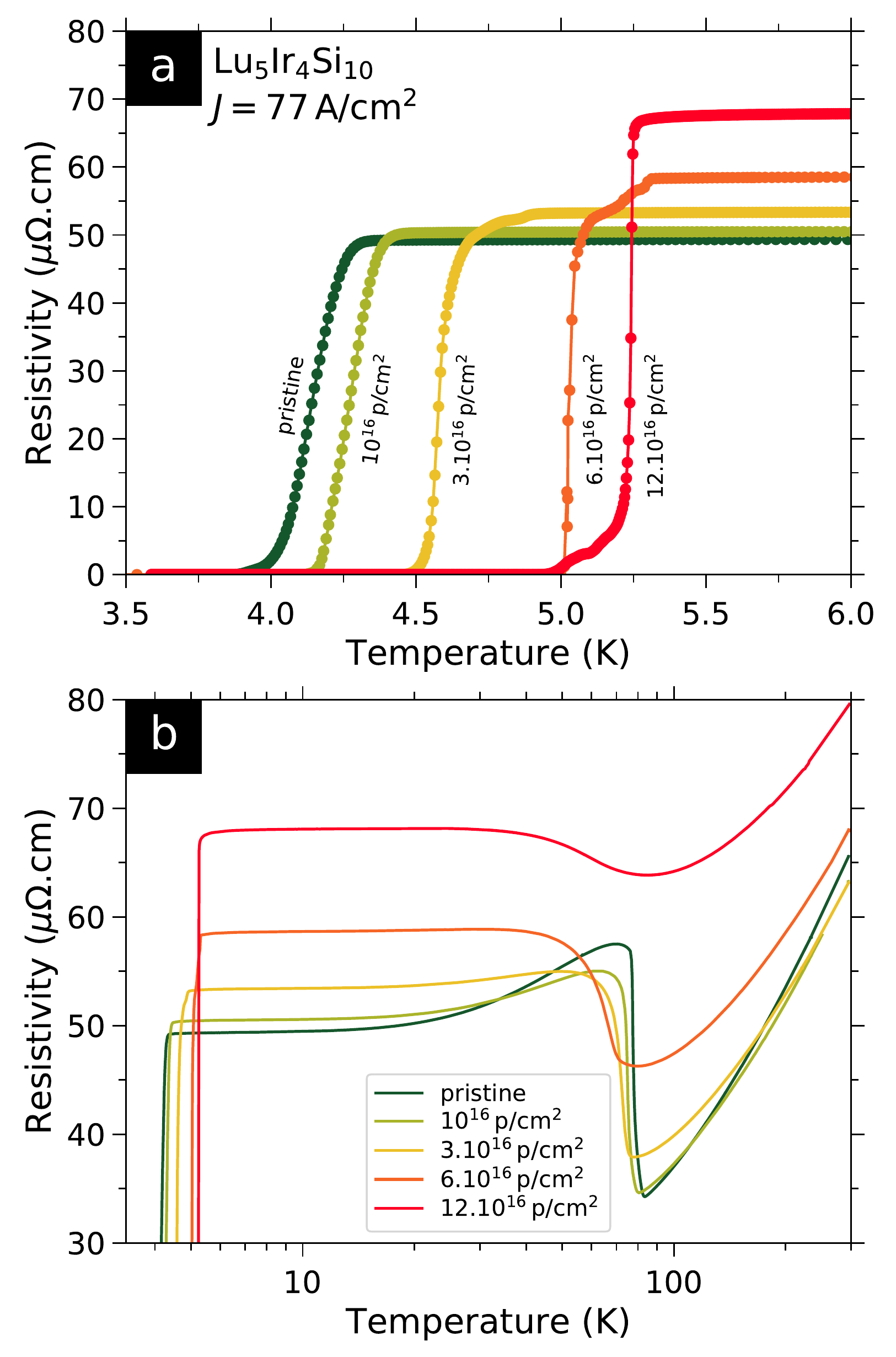}
\caption{\label{fig:RT_aTc_bTcdw}\textbf{Temperature dependence of the c-axis resistivity of Lu$_5$Ir$_4$Si$_{10}$ for increasing irradiation doses.}
\textbf{(a)} The superconducting transition shifts to \emph{higher} temperature after irradiation, in contrast with expected behavior.
The transition width (15 - 85 \%) surprisingly \emph{decreases} with irradiation, evidencing uniform irradiation damage (see text), even though there appear shoulders of unknown origin near the top / bottom of the superconducting transitions at high doses.
\textbf{(b)} The large increase in resistivity below 80\,K is caused by a CDW which gaps density of states at the Fermi level and increases electronic scattering.
Contrary to \Tc, the CDW transition temperature shifts to \emph{lower} temperature, the transition width \emph{increases}, and the amplitude of the increase is reduced for increasing irradiation dose.}
\end{figure}
\section{Disorder increases \Tc\ and decreases \TCDW}\label{disorderTc}
Fig.~\ref{fig:RT_aTc_bTcdw} shows the evolution of the superconducting (panel a) and of the CDW transition (panel b) with increasing irradiation dose.  We define \Tc\ as the point where the resistivity dropped to 50\% (midpoint). Fig.~\ref{fig:RT_aTc_bTcdw}a reveals a clear increase of \Tc\ with irradiation, from 4.15\,K in the pristine state to 5.25\,K at the highest irradiation dose.  Even though there appear shoulders of unknown origin near the top / bottom of the superconducting transitions at high doses, the width of the main part of the transition (15 – 85\%) surprisingly decreases upon irradiation.

Fig.~\ref{fig:RT_aTc_bTcdw}.b, shows an overview of the c-axis resistivity of Lu$_5$Ir$_4$Si$_{10}$  in semi-log scale. The transition to the CDW phase at low temperature appears as a large increase of resistivity below $T_\mathrm{CDW} \approx 77$\,K, as previously observed\cite{Shelton1986LIS_P_rho_Cp}.
We define \TCDW\ as the midpoint (50\%) of this increase in resistivity.
As the irradiation dose increases, \TCDW\ shifts toward lower temperature and the amplitude of the increase in resistivity is reduced.
Contrary to the superconducting transition, the width of the CDW transition strongly increases with irradiation dose.
% hysteresis
The CDW transition also has an hysteresis of approximately 1\,K, as was previously observed\cite{OpagisteLeroux2010LISwhiskers,Leroux2013LIS_SCprop} and recently studied in details\cite{SAINTPAULJPCS2020}. We find that this hysteresis survives up to the highest irradiation dose. However, after irradiation the hysteresis occurs only below \TCDW\ (midpoint), whereas, in the pristine state, the hysteresis extends up to the onset of the transition (83.5\,K).

\begin{figure}
\includegraphics[width=0.5\textwidth]{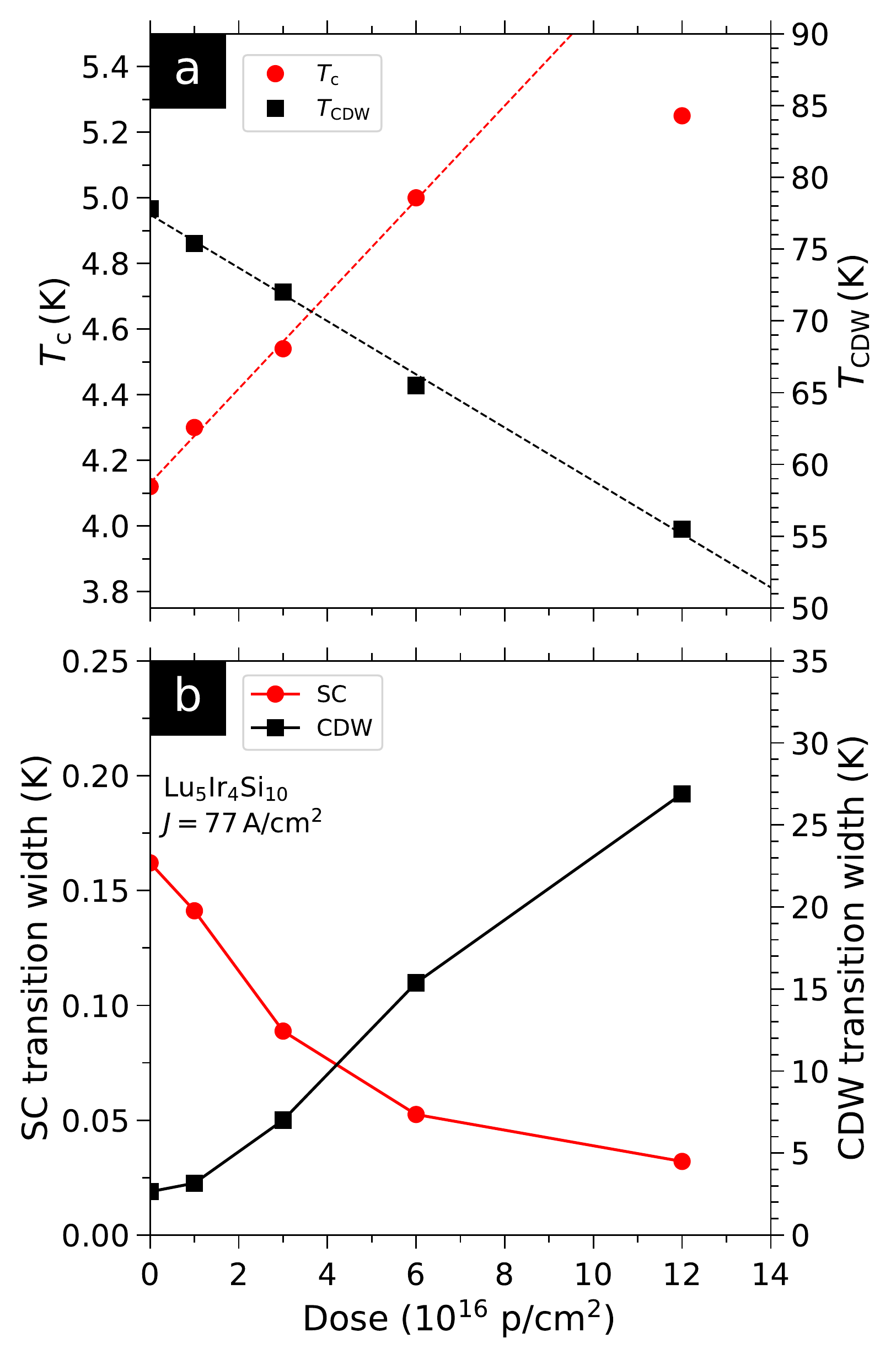}
\caption{\label{fig:Tc_Tcdw_vsrho0}
\textbf{Superconducting and CDW transitions as a function of irradiation dose.}
\textbf{(a)} \Tc\ and \TCDW\ vary linearly up to high irradiation doses.
\TCDW\ decreases at a constant rate of -1.85\,K/10$^{16}$ p/cm$^2$, whereas \Tc\ increases at a rate of +0.14\,K/10$^{16}$ p/cm$^2$, which appears to saturate at the highest irradiation dose.
\textbf{(b)} For both transitions we define the width using a 15\%\,--\,85\% criterion (for the CDW: between the min/max resistivity above/below \TCDW, respectively).
As irradiation dose increases, the superconducting transition width is reduced whereas the CDW transition width increases, evidencing the different mechanisms through which disorder affects them (see text).
}
\end{figure}

We note that Matthiessen's rule seems to fail both above and below \TCDW\ in Fig.~\ref{fig:RT_aTc_bTcdw}.b. But one needs to be very careful when applying Matthiessen's rule in a CDW system. In the CDW phase, the density of states (DOS) at the Fermi level decreases (gapped charge carriers) but there are also fewer electrons to scatter off, meanwhile the CDW will also contribute to electronic scattering. So, the resistivity can go either way at the CDW transition.
A blatant example of this mercurial behavior are the 1T and 2H polytypes of the dichalcogenide compound TaS$_2$: the resistivity increases in the CDW phase of 1T- TaS$_2$ (dominated by DOS effects), while the resistivity decreases in the CDW phase of 2H- TaS$_2$ (dominated by electron scattering effects).
\textcolor{black}{Note that a similar effect of increase/decrease in resistivity can also occur at a SDW transition as observed in BaFe$_2$(As$_{1-x}$P$_x$)$_2$: irradiating the sample with electrons makes the transition split into an upturn and subsequent downturn upon cooling\cite{Mizukami2017JPSJ}.}
Besides, the CDW is commensurate in Lu$_5$Ir$_4$Si$_{10}$, so that there is also a Fermi Surface reconstruction, and it is a system with tens of electronic bands, eleven of which cross the Fermi level.\cite{Mansart2012LIS_femtosec_and_DFT}
Finally, a short range CDW order can persist above the damage level erasing the trace of transition in the resistivity versus temperature curve, as evidenced in NbSe$_2$ in Ref.\citenum{Cho2018}, and short range CDW correlations can also persist above \TCDW\ (e.g. CDW satellite peaks already appear in x-ray below 85 K in Ref.\citenum{Becker1999}). Such a short range CDW order will not lead to Fermi surface reconstruction but will still contribute to transport properties.

% variations of Tc
The simultaneous variations of \TCDW\ and \Tc\ are summarized in Fig.~\ref{fig:Tc_Tcdw_vsrho0} as a function of irradiation dose. \Tc\ increases almost linearly at a rate of +0.14\,K/10$^{16}$ p/cm$^2$ (or 0.093\,K/$\mu\Omega$.cm) and starts saturating after the last irradiation.
\TCDW\ decreases linearly in the whole range of irradiation doses at a rate of -1.85\,K/10$^{16}$ p/cm$^2$ (or -1.18\,K/$\mu\Omega$.cm).

According to Anderson's theorem\cite{Anderson1959}, in an isotropic s-wave superconductor small concentration of non-magnetic defects should not affect \Tc\ while magnetic defects should be pair-breaking and reduce \Tc.
Generally, the effect of pair-breaking scattering is described by Abrikosov-Gorkov theory\cite{abrikosov1961aa, Openov98}. In this theory, \Tc\ is found to always decrease\cite{AlbenquePRLYBCO7,Albenque,AlloulHirschfeldRMPdefects}.
We thus conclude that the increase of \Tc\ we observed, cannot be explained by the standard effects of disorder on a superconductor. 

Rather, such an increase of \Tc\ with irradiation dose arises naturally from a competition scenario betwen the CDW and superconductivity, if irradiation suppresses the density wave more than superconductivity\cite{Mishra2015SDW,Fernandes2012Tcupwithdisorder, Grest82, Psaltakis84}. We recently demonstrated such an increase of \Tc\ via competition with CDW using irradiations in the d-wave cuprate superconductor \LBCO\cite{LerouxPNAS2019}.
This has also been evidenced in the dichalcogenides superconductors using irradiation induced disorder\cite{Mutka1983thesis,Mutka1983PRB_CDW_SC_irrad,Cho2018} and substitution disorder\cite{Chatterjee2015}. Both types of disorder strongly suppress CDW, either via real-space phase fluctuations\cite{McMillan1975PRB,Mutka1981PRB1TTaS2} (domains) or by pair-breaking\cite{Stiles1976NbSe2}.
A competition scenario was also proposed for \LIS\ based on pressure\cite{Shelton1986LIS_P_rho_Cp} and doping\cite{SinghPRB2005,Leroux2013LIS_SCprop} studies.
The increase of \Tc\ that we observe upon irradiation, is therefore definitive evidence that the CDW is competing with superconductivity in \LIS.

Quantitatively, Lu$_5$Ir$_4$Si$_{10}$ is a weak coupling limit s-wave superconductor ($\Delta C/ \gamma T_\mathrm{c}$ = 1.41\cite{Shelton1986LIS_P_rho_Cp}, close to 1.43), so that the electronic density of states released by the CDW should yield an exponential increase of \Tc\ following the standard formula for a BCS superconductor: $T_\mathrm{c} = \alpha \, \theta_D \,\exp\left({-\frac{1}{N(E_\mathrm{F})V}}\right)$ where $\alpha\approx1.14$ in the weak coupling limit, $\theta_D=[315-366]$\,K is the Debye temperature\cite{Leroux2013LIS_SCprop,Shelton1986LIS_P_rho_Cp}, $N(E_F)$ is the density of states at the Fermi level involved in Cooper pairs and V is the attractive potential between the electrons of the pair.

\begin{figure}
\includegraphics[width=0.5\textwidth]{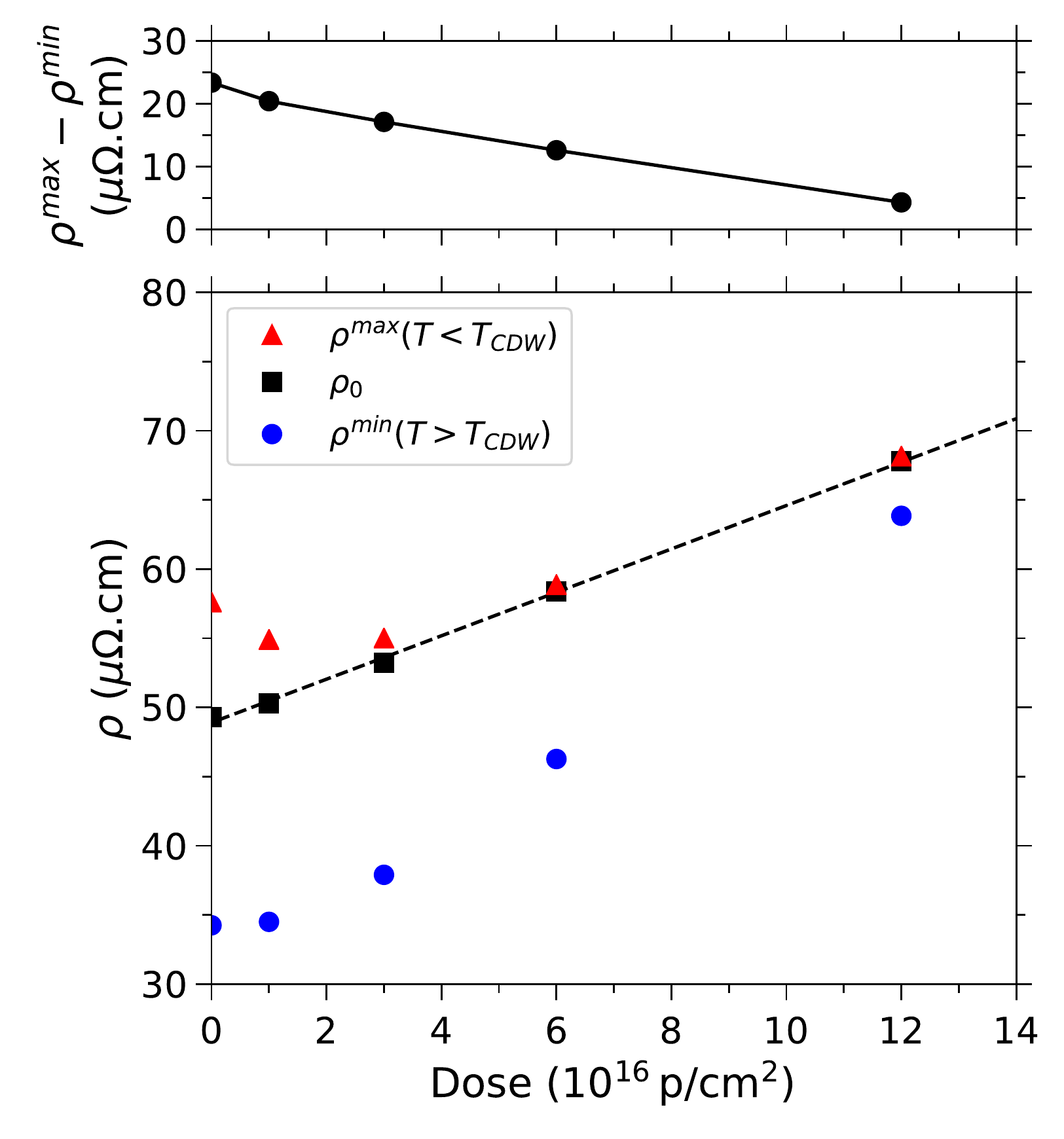}
\caption{\label{fig:rho0_vsdose}
\textbf{Irradiation dose dependence of the residual resistivity  $\rho_0$ and the resistivity jump at \TCDW.} We define the amplitude of the jump as the difference $\rho^{max}-\rho^{min}$ between the maximum and minimum resistivity below and above \TCDW, respectively (see Fig.~\ref{fig:RT_aTc_bTcdw}.b). $\rho_0$ increases linearly with dose at a rate of 1.57\,$\mu\Omega$.cm/$10^{16}$\,p/cm$^2$. No saturation of defects creation is observed up to $12\times10^{16}$\,p/cm$^2$.
}
\end{figure}

The evolution of the residual resistivity as a function of irradiation dose is presented in Fig.~\ref{fig:rho0_vsdose}.
Before irradiation, the residual resistivity is $\rho_0 =49.3\,\mu\Omega$.cm and the residual resistivity ratio (RRR) is $\rho_\mathrm{295\,K}/\rho_0=1.3$, in-line with previous studies\cite{Shelton1986LIS_P_rho_Cp,SinghPRB2005,OpagisteLeroux2010LISwhiskers}.
After irradiation, we find that the residual resistivity increases linearly at a rate of 1.57\,$\mu\Omega$.cm/$10^{16}$\,p/cm$^2$ without saturation up to $12\times10^{16}$\,p/cm$^2$.
Such a linear increase is what is typically expected in metals following the "unitary limit"\cite{SunkoDelafossiteIrradPRX}, but this was not \textit{a priori} obvious in \LIS\ because of the CDW.
Indeed, on the one hand, irradiation suppresses the CDW, which increases the density of states at the Fermi level and should reduce the residual resistivity.
But on the other hand, irradiation increases the number of defects and reduces the size of CDW domains, both of which should raise electronic scattering and increase the residual resistivity.
Here, we can at least conclude that the increased scattering more than compensates the increased density of states, as the residual resistivity increases overall.
We also find that  $\rho^{max}-\rho^{min}$, the amplitude of the jump in resistivity at \TCDW, is reduced after irradiation (see Fig.~\ref{fig:rho0_vsdose}). Again, a natural explanation for this reduction would be that after the CDW is suppressed by disorder, it does not gap as much electronic density of states.

%discrepancies in the literature
We also note that there appear to be slight discrepancies in the literature values of \Tc\ and \TCDW\ in pristine single crystals\cite{Becker1999,Shelton1986LIS_P_rho_Cp,Yang91,SinghPRB2005,JungMiglioriPressure2003}: ranging from \Tc\ = 3.8\,K and \TCDW\ = 79\,K, to \Tc\ = 3.9\,K and \TCDW\ = 83\,K. This could be due to different criteria to determine the transition temperatures, for instance using the onset of the jump in resistivity versus the midpoint to define \TCDW . This onset is also at 83 K in our work and we expect it to be less sensitive to disorder as it relates to short range CDW fluctuations. But based on our present results, the spread in \Tc\ and \TCDW\ values in the literature could simply reflects the slightly different levels of disorder in as-grown crystals.

%%%%%%%%%%%%%%%%%%%%%%%%%%%%%%%%%%%%%%%%%%%%%%%%%%%%%
%%%%%%%%%%%%%%%%%%%%%%%%%%%%%%%%%%%%%%%%%%%%%%%%%%%%%
%%%%%%%%%%%%%%%%%%%%%%%%%%%%%%%%%%%%%%%%%%%%%%%%%%%%%
% part VI
\section{Increase of $H_\mathrm{c,2}$ with disorder}\label{magfield}

\begin{figure}
\includegraphics[width=0.9\textwidth]{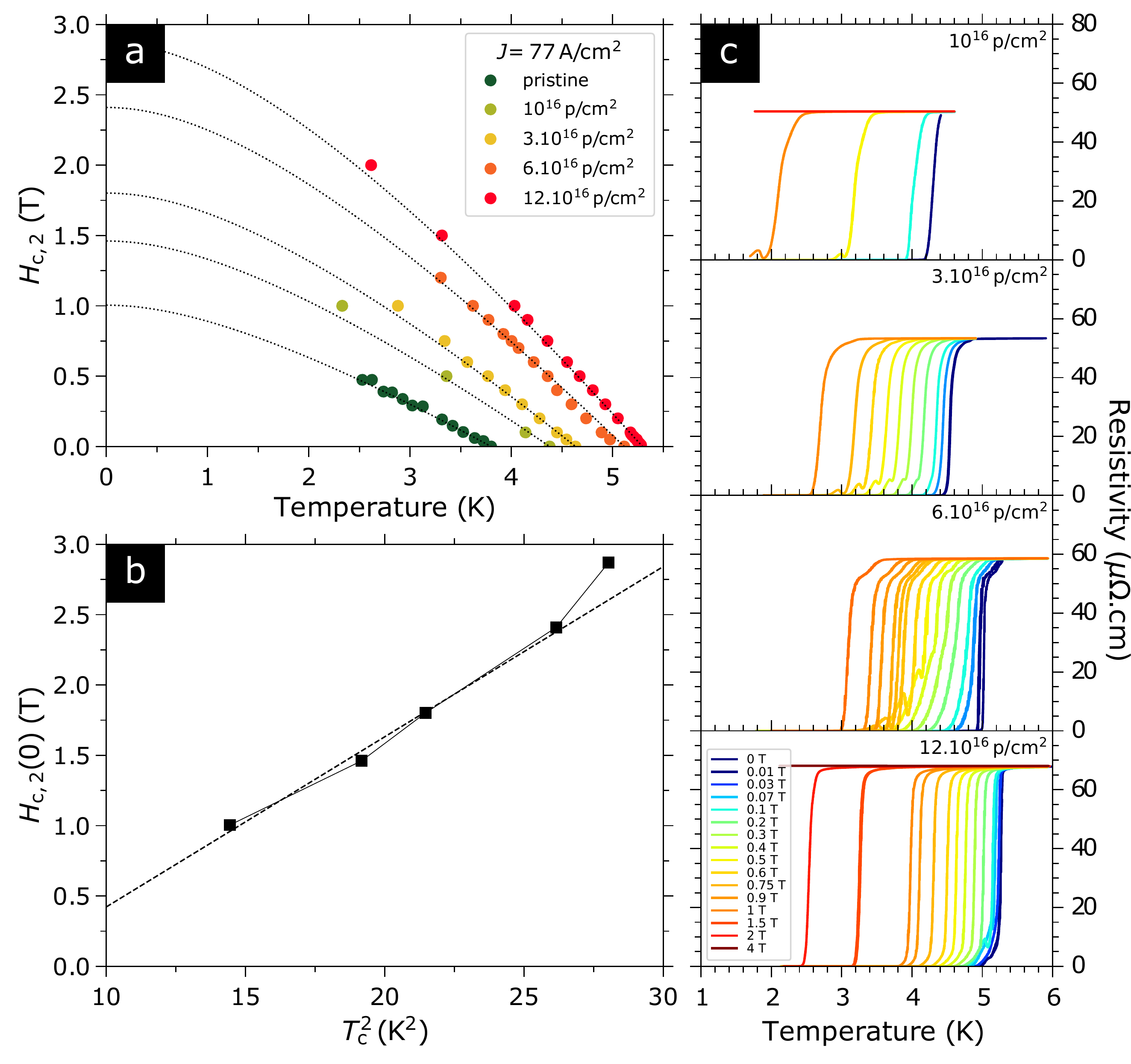}
\caption{\label{fig:aHc2_vsT_bHc20vsT2}\textbf{$H_\mathrm{c,2}$ vs irradiation dose.} \textbf{(a)} Superconductivity second critical field $H_\mathrm{c,2}$ for H in-plane as a function of temperature and irradiation dose. Lines are theoretical curves from Werthamer-Helfand-Hohenberg-Maki (WHHM) theory\cite{WHHM_PR_1966} with $\alpha = 0.21$ and $\lambda_\mathrm{SO}=0$. Data in the pristine state is taken from Ref.~\citenum{Shelton1986LIS_P_rho_Cp}. \textbf{(b)} Superconductivity second critical field at zero temperature $H_\mathrm{c,2}(0)$, extrapolated from the WHHM curves in a), appears to increase linearly with $T_\mathrm{c}^2$ at a rate of 0.12 T/K$^2$ (see text). \textbf{(c)} Detailed magnetic field dependence (H//ab) of the c-axis resistivity of Lu$_5$Ir$_4$Si$_{10}$ at the superconducting transition, using a current density of 77 A/cm$^2$ and for increasing irradiation doses (top to bottom). The superconducting transition does not broaden, even after repeated irradiations and in a magnetic field. A vortex lattice peak effect is also visible in the superconducting transition (see also Fig.~\ref{fig:RT_vsH_vsJ_log} in appendix).}
\end{figure}

As expected from the increase of \Tc\, we find an increase of the in-plane upper critical field ($H_\mathrm{c,2}$) for increasing irradiation dose. Our measurements of $H_\mathrm{c,2}$ are reported Fig.\ref{fig:aHc2_vsT_bHc20vsT2}a, as a function of temperature for several irradiation doses, and where we define $H_\mathrm{c,2}$ as the point where the resistivity drops 10\% below the residual resistivity value $\rho_0$.
Fig.\ref{fig:aHc2_vsT_bHc20vsT2}c shows the detailed magnetic field dependence of the R(T) curves at the superconducting transition. We note that, notwithstanding the peak effect related to vortex lattice softening, the R(T) curves are shifted downward as the field increases, with the same shape and without broadening, even after repeated irradiations and in large magnetic fields.
Our data is in good agreement with published $H_\mathrm{c,2}(T)$ data in non irradiated Lu$_5$Ir$_4$Si$_{10}$\cite{Shelton1986LIS_P_rho_Cp, JAISWAL2002142}. 
We find that $H_\mathrm{c,2}(T)$ follows the Werthamer-Helfand-Hohenberg-Maki (WHHM) scaling\cite{WHHM_PR_1966} at all irradiation doses.
Using the WHHM scaling, Ref.~\citenum{Shelton1986LIS_P_rho_Cp} found best fit parameters values $\alpha = 0.21$ for the Maki parameter and $\lambda_\mathrm{SO}=9.0$ for the spin-orbit coupling.
However, due to the small value $\alpha=0.21$, fits are essentially insensitive to the choice of the spin-orbit coupling $\lambda_\mathrm{SO}$, so that we can adopt in the following $\lambda_\mathrm{SO}=0$ and extrapolate the value of $H_\mathrm{c,2}(0)$ with little uncertainties (see appendix A for details).

In Fig.~\ref{fig:aHc2_vsT_bHc20vsT2}b we find that $H_\mathrm{c,2}(0)$ is in good agreement with a $T_\mathrm{c}^2$ dependence.
In a usual s-wave isotropic superconductor in the dirty limit, $H_\mathrm{c,2}(0)$ should scale with \Tc. Indeed, the upper critical field is equal to $\frac{\phi_0}{2\pi\xi^2_0}$ and in the dirty limit the coherence length $\xi_0$ is renormalized to $\xi_{0,d} = \sqrt{\xi_0 \bar{l_0}}$, where $\bar{l_0} = v_\mathrm{F} \tau_0$ and $\xi_0=\frac{\hbar v_F}{1.76 \pi k_BT_c}$ in the weak-coupling BCS theory, yielding : 
\begin{equation}
\label{eq:Hc20dirty}
H_\mathrm{c,2}(0) = \frac{\phi_0}{2\pi \xi_{0,d}^2} \approx \frac{\phi_0  1.76\,k_\mathrm{B}}{2 \hbar v_\mathrm{F}^2}\,\frac{1}{\tau_0} T_\mathrm{c}
\end{equation}
So usually, in the dirty limit $H_\mathrm{c,2}(0) \propto T_\mathrm{c}$, however here $H_\mathrm{c,2}(0) \propto T_\mathrm{c}^2$. This unusual scaling can be easily explained by the fact that $1/\tau_0$ varies linearly with \Tc\ due to the competition with the CDW,
namely: the prefactor $\frac{\phi_0  1.76\,k_\mathrm{B}}{2 \hbar v_\mathrm{F}^2}$ is independent of irradiation dose, whereas  $1/\tau_0$ is usually proportional to the irradiation dose for uniform non-overlapping defects in metals\cite{SunkoDelafossiteIrradPRX}, and in this compound, empirically, \Tc\ increases linearly with the irradiation dose (see Fig.\ref{fig:Tc_Tcdw_vsrho0}).

%%%%%%%%%%%%%%%%%%%%%%%%%%%%%%%%%%%%%%%%%%%%%%%%%%%%%
%%%%%%%%%%%%%%%%%%%%%%%%%%%%%%%%%%%%%%%%%%%%%%%%%%%%%
%%%%%%%%%%%%%%%%%%%%%%%%%%%%%%%%%%%%%%%%%%%%%%%%%%%%%
% part IV
\section{Pressure, doping and irradiation: progressive versus binary increase of \Tc}\label{Tcpresdopdis}
Interestingly, the competition scenario between CDW and superconductivity still requires clarification in \LIS.
Indeed, pressure studies\cite{Shelton1986LIS_P_rho_Cp,JungMiglioriPressure2003} show a \emph{binary} effect on \Tc: below 2\,GPa, \Tc\ is 4\,K and constant, whereas above 2\,GPa, \Tc\ is 9\,K and constant; in contrast doping studies\cite{Yang91,SinghPRB2005,Leroux2013LIS_SCprop} show a \emph{progressive} increase of \Tc\ from 4\,K to 6\,K. This is all the more surprising as in both cases the CDW is progressively suppressed. Hence, naively, shouldn't one expect that \Tc\ also increases progressively with pressure ?
We argue that this can be explained by the different mechanisms through which the CDW is suppressed when using pressure, doping and irradiation.

% Tc Tcdw width
The variations as a function of irradiation dose of the width (15 - 85 \%) of the superconducting and CDW transitions are summarized in Fig.~\ref{fig:Tc_Tcdw_vsrho0}.b.
As can be seen, the width of the superconducting transition decreases with irradiation dose.
In general, for the superconducting pairs in a s-wave superconductor, disorder only acts via the pair-breaking mechanism from magnetic defects.
%%%NEW
The defects introduced by proton irradiation are non-magnetic in nature. Therefore, they are not pair-breaking for a s-wave superconductor like \LIS. The Anderson theorem remains valid even in the presence of cascade defects, as they are not magnetic defects.\cite{Anderson1959}
In addition, the suppression mechanism by real-space phase fluctuations from disorder is usually irrelevant for superconductivity unless the superconducting condensate is modulated in real-space, that is Cooper pairs have a non-zero center of mass momentum as in the FFLO, stripes or pair-density wave superconducting states\cite{FradkinTranquada2015RoMP,DavisPNAS2013}. The question remains open in \LIS\ whether the superconducting condensate is modulated in the CDW phase (beyond a trivial spatial segregation), but we do not think this is highly likely as it is a rather standard BCS s-wave compound. STM studies as in Bi$_2$Sr$_2$CaCu$_2$O$_8$\cite{EdkinsScience2019} could shed light on this interesting open question. So, except for the elusive non-zero-momentum Cooper pairs states, in a standard s-wave superconductor no change of the superconducting transition width is expected with disorder at first. However, there could still be a ``trivial'' broadening of the superconducting transition via the competition with the CDW, that is: a spatially inhomogeneous CDW will yield a spatially inhomogeneous superconducting order parameter (stronger where the CDW is weaker), hence a broader \Tc.
 
Thus, here the decreased superconducting-transition width after irradiation shows that the irradiation damage caused by proton irradiation is very uniform, to the point that disorder in the sample is actually more uniform after irradiation.

Let us now turn to how disorder influences the CDW. By contrast to the superconducting transition, as shown in Fig.~\ref{fig:Tc_Tcdw_vsrho0}, the width of the CDW transition significantly increases with irradiation dose,
and both \TCDW\ and the jump in resistivity decreases with increasing disorder.
These are strong indications that the CDW is suppressed by real-space phase fluctuations.

While the almost linear variation of residual resistivity with irradiation dose points to the main contribution arising from scattering in the unitary limit, 4-MeV proton irradiation produces a complex morphology of defects: beyond vacancy-interstitial pairs (atomic point defects), cascade-type clusters of defects are also generated. These clusters are typically a few nanometers in diameter, and, because of this larger size, they are expected to be less pair-breakers for superconductivity than point defects \textcolor{black}{as bigger defects should scatter in a smaller volume of reciprocal space and be closer to or larger than the coherence length. There is also experimental evidence for this lesser pair breaking effect: an extreme example is commercial superconducting tapes based on cuprates. Despite the huge density of tiny nanoparticles in these tapes, \Tc\ is almost the same as in the clean bulk cuprate and proton irradiation also does not affect \Tc\ much in these tapes.\cite{Jia2013protonirrad}.
Also, in iron-based superconductors, it was found that the rate of depression of \Tc\ per unit of residual resistivity increase is twice higher when using electron irradiation\cite{Mizukami2014NatComirradFeSC} (producing solely point defects) rather than proton irradiation\cite{Smylie2016PRB}, suggesting that cascade defects could have less pair breaking effect on superconductivity too. 
We note that in the case of proton irradiation the scattering process from combined point defects and cascade-type clusters of defects has an undetermined parameter: the ratio between those two channels. In superconductors where non-magnetic defects are pair-breakers, such as those with s+- or d-wave gap, this complex damage morphology can impede conclusive statements on pairing mechanism. For instance, a decrease of \Tc\ followed by a saturation of \Tc\ for increasing proton irradiation dose has been attributed to a crossover in pairing mechanism from s+- to s++\cite{GhigoPRL2018}, but this effect could not be reproduced in electron irradiation studies\cite{Cho2018PnictideElecIrradReview}.}

In \LIS\, the cascade defects are expected to not be pair-breaking at all as we know that non-magnetic defects do not decrease \Tc\ in an isotropic s-wave superconductor (which is the case here until proven otherwise). However, nonmagnetic defects irrespective of their sizes do suppress the CDW order, which is consistent with the Imry-Ma theorem\cite{ImryMaTheorem}, and some earlier studies.\cite{DisorderOnCDWTheory,DisorderOnCDWMeanField} Therefore, the cascade defects have pair-breaking effect only on the CDW phase. In addition, we expect the cascade-type defects produce strong pinning of the CDW and contribute to the broadening of the CDW transition.
Therefore, for the study of the coexistence of superconductivity and CDW, it could be seen as a benefit of proton irradiation that it provides these few nanometers wide cascade-type defects. These cascades should indeed mostly suppress the CDW while not affecting much superconductivity \textcolor{black}{(even if of the d-wave type\cite{Jia2013protonirrad,LerouxPNAS2019})}, thus enabling to evidence the competition or synergy between the two more clearly.

%Indeed,
Two different mechanisms have been proposed for CDW suppression by disorder: (i) a pair-breaking mechanism\cite{JohannesMazin,Mutka1983thesis}where disorder increases the scattering rate, which induces a broadening of the Fermi function and reduces the peak in electronic susceptibility. This process reduces the jump in resistivity at the CDW transition, but it does not affect the macroscopic coherence of the CDW and produces a uniform global reduction of the CDW. Thus this mechanism cannot account for the broadening of the CDW transition. (ii) a real-space phase fluctuations mechanism where disorder pins the phase of the periodic spatial modulation of the electronic density and associated lattice distortion. This breaks up the CDW into small domains, which broadens the transition(Ref.~\citenum{VojtaCupratesReview2009} §2.7.2).
Thus, our data indicate that real-space fluctuations are clearly contributing to the suppression of the CDW by irradiation in this study.
In doping studies\cite{Yang91,SinghPRB2005} in which Sc (Co) were introduced on the Lu (Ir)-sites, respectively, pronounced broadening and suppression of the CDW was observed, in analogy to the results presented here, suggesting that real-space phase fluctuations are suppressing the CDW in doping studies as well. However, in doping studies, additional effects may arise from doping-induced changes of the Fermi surface.

Conversely, in pressure studies, the mechanism by which the CDW is suppressed must be different as there is no change in the number of defects, hence no pair-breaking nor phase fluctuations. This is also quite strikingly evidenced experimentally\cite{Shelton1986LIS_P_rho_Cp,JungMiglioriPressure2003}: even though \TCDW\ decreases by up to a factor of 10, there is no significant change in the CDW transition width under pressure.
A natural explanation for this is that pressure stiffens the elastic constants of the crystal, which makes the periodic lattice distortion less favorable energetically and reduces \TCDW. This suppression-by-elastic-stiffening mechanism follows from the standard CDW stability criterion of Chan and Heine\cite{ChanHeine}.
As this mechanism is global in essence, it explains why the CDW transition width remains sharp and constant even though \TCDW\ diminishes.
Finally, as \Tc\ remains essentially constant up to 2\,GPa, it also means that the density of states at the Fermi level is essentially constant with pressure in the CDW phase. To first order, the elastic stiffening modifies only the temperature at which the Chan and Heine criterion is met, without affecting the density of electronic states involved. 

Thus, we can now reconcile the effects of pressure, doping and irradiation on the CDW and superconductivity, by considering the differences between the three mechanisms of CDW suppression: pair-breaking, real-space phase fluctuations and elastic constants stiffening.
The main effect of pressure on the CDW is to reduce \TCDW\ via elastic constants stiffening, but both the CDW-transition width and the amount of density of states gapped by the CDW remains constant. Therefore \Tc\ increases in a binary way: as pressure increases \Tc\ is essentially constant and \TCDW\ decreases, until $T_{\mathrm{CDW}} < T_{\mathrm{c}}$, at which point \Tc\ jumps to another constant value.
By contrast, with doping or irradiation there are both phase fluctuations and pair-breaking which not only reduces \TCDW, but also reduces the amount of density of states gapped by the CDW. Thus \Tc\ increases in a progressive way: doping/irradiation frees density of states that was gapped by the CDW, which then raises \Tc\cite{BilbroMcMillan1976} (as superconductivity is not subject to the same strong pair-breaking or phase fluctuations effects of disorder).

We note here, that CDW and superconductivity cannot live \emph{independently} as the CDW gap can develop only on some parts of the Fermi surface, but the superconducting gap always opens over the entire Fermi Surface. So, in this sense, CDW and superconductivity are always competing for electrons at the Fermi level, albeit somewhat trivially. An even stronger evidence for competition would be a reduction of the CDW diffraction peak upon entering the superconducting state, as in x-ray studies of YBa$_2$Cu$_3$O$_{6.67}$\cite{Chang2012YBCOcompetingCDW}. To our knowledge, this is an open question that has not been settled experimentally in \LIS\ and would greatly benefit from low temperature x-ray studies. Likewise, if the CDW modulation survives in the superconducting state, we would expect some comodulation of the superconducting and CDW order parameters, if only around defects and close to vortex cores, which an STM study might be able to evidence, if proper surfaces of \LIS\ can be produced.

\section{Conclusion}
We showed that irradiation induced disorder enhances the superconducting critical temperature \Tc\ and $H_\mathrm{c2}$ while it suppresses the CDW in \LIS.
We showed how this increase of \Tc\ cannot be accounted for by the expected effect of disorder, and instead stems from the increase of density of states at the Fermi level.
Our results thus make a very compelling case that superconductivity and CDW are competing for electronic density of states at the Fermi level in \LIS.
Owing to its prototypical, 1D, Peierls type CDW and the s-wave, weak-coupling nature of its superconductivity, \LIS\ thus provides a platform from which to understand the more complex cases of density waves and superconductivity coexistence in heavy fermions, Fe-based or cuprates superconductors.
Recently\cite{HirschfeldPRL2018} it was shown that, in an unconventional superconductor mediated by spin-fluctuation, very inhomogeneous conditions may result in an increase of \Tc. However we do not think this latter case is relevant to the standard s-wave superconductor studied here.
Also very recently, it was found that, while the effect of disorder on the dichalcogenide NbSe$_2$ in bulk form is explained in terms of CDW-superconductivity competition and synergy\cite{Cho2018}, in monolayer NbSe$_2$ a much larger \Tc\ dome was discovered as a function of disorder and this has been proposed to be due to the wavefunction multifractality in a 2D monolayer system\cite{Zhao2019,rubioverd2018visualization}. Thus, disorder as a tuning parameter is finding relevance not only for the study of bulk superconductors with density waves but also for 2D materials such as monolayer transition metal dichalcogenides.
%Interestingly, from a technological perspective, our results also unlock the possibility of direct-writing superconducting detectors and Josephson junctions by localized irradiation via a mask or a helium FIB, which could support the current effort in Quantum Information Science based on SQUID technology.

% If you have acknowledgments, this puts in the proper section head.
\begin{acknowledgments}
M.L. acknowledges fruitful discussions with S.~Eley, F.~Ronning, C.~Proust and P. Monceau. The experimental study at Argonne National Laboratory was supported by the US Department of Energy (DOE), Office of Science, Materials Sciences and Engineering Division. V.M. was supported by the Center for Emergent Superconductivity, an Energy Frontier Research Center, funded by the US DOE, Office of Science, Office of Basic Energy Sciences. Proton irradiation was performed at Western Michigan University. CO and PR work are supported by the ANR-DFG grant ANR-18-CE92-0014-03 "Aperiodic". The samples used in this study were grown at the N\'eel Institute by C.~Opagiste. 
\end{acknowledgments}

% Create the reference section using BibTeX:
\bibliography{biblio}

\section{Appendices}
\appendix
\section{WHHM Scaling}
\begin{figure}
\includegraphics[width=\linewidth]{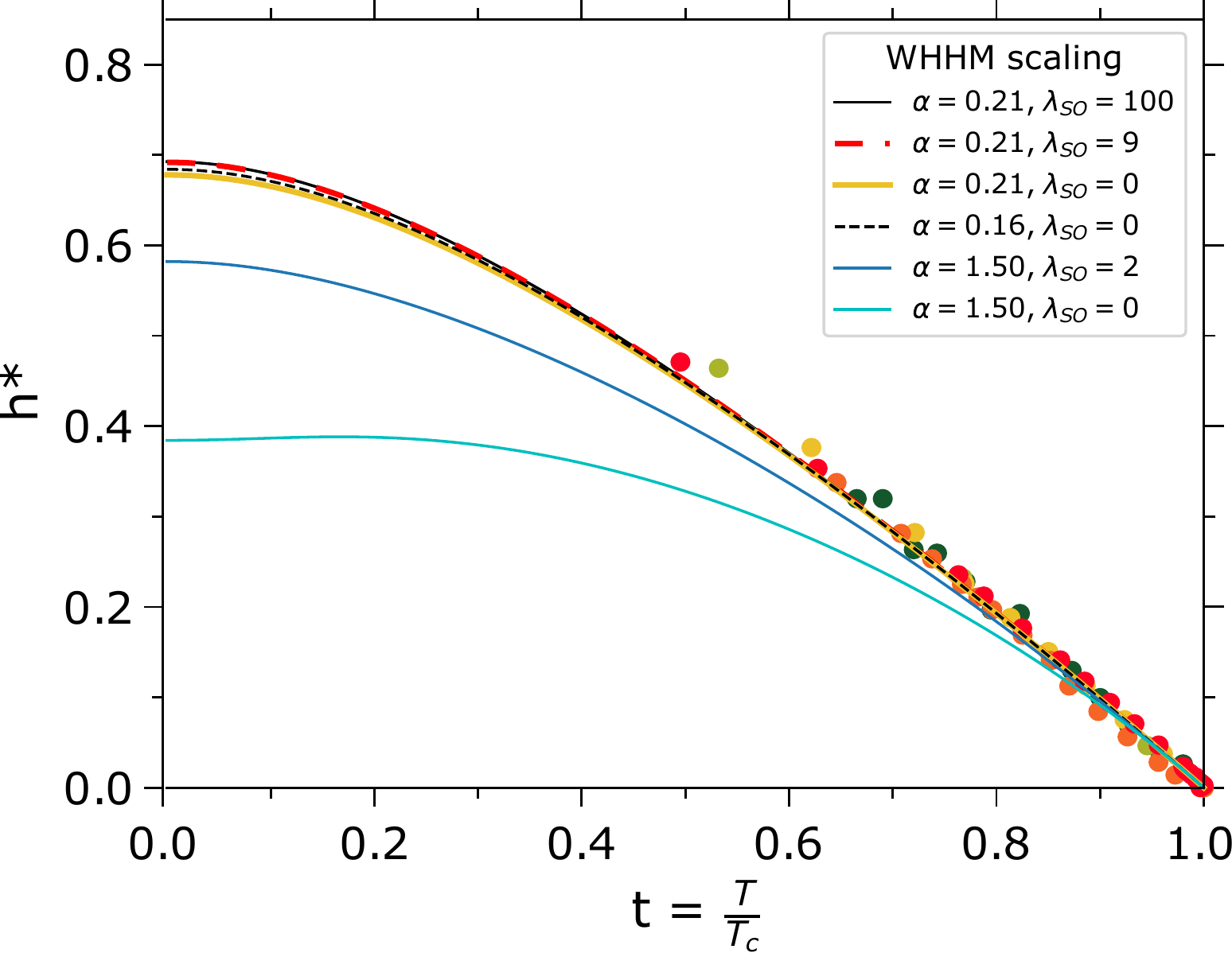}
\caption{\label{fig:aHc2_vsT_bWHHM}\textbf{$H_\mathrm{c,2}$ and WHHM scaling.} Superconductivity reduced second critical field h* = $H_\mathrm{c,2} / \left(-\mathrm{d}H_\mathrm{c,2}/\mathrm{d}t\right)_{_{t=1}}$ as a function of the reduced temperature t = $T/T_c$, for all irradiation doses. No significant changes occur with irradiation in this reduced plot. The lines are theoretical curves from WHHM theory. Data in the pristine state was taken from Ref.~\citenum{Shelton1986LIS_P_rho_Cp}.}
\end{figure}

To support the relevance of the WHHM theory in \LIS, despite its complex Fermi Surface\cite{Mansart2012LIS_femtosec_and_DFT}, we verify that the $\alpha$ values estimated in two independent ways are close, following the recommendation of Ref.~\citenum{WHHM_PR_1966}. First, using the normal state values just above \Tc : 
$ \alpha = \frac{3e^2\hbar\gamma\rho_0}{2m\pi^2k_\mathrm{B}^2} \approx 0.16 \, \mathrm{(SI\ units)}$, 
with $\gamma=119.1$\,J/m$^3$/K$^2$ (the volumic Sommerfeld coefficient from $\gamma=23.42\,\mathrm{mJ/mol/K}^2$ in Ref.\citenum{Shelton1986LIS_P_rho_Cp}) and $\rho_0=56.2\,\mu\Omega$.cm (also from Ref.\citenum{Shelton1986LIS_P_rho_Cp}).
This result is close to the value deduced from the slope of $H_\mathrm{c,2}$ near \Tc (in T/K):
$\alpha = 0.52758\times -\left(\frac{dH_\mathrm{c,2}}{dT}\right)_{_{T_c}} \approx 0.21 \, \mathrm{(SI\ units)}$.
In Fig.~\ref{fig:aHc2_vsT_bWHHM}, we show that because of the small value $\alpha=0.21$, the WHHM scaling is essentially insensitive to the choice of the spin-orbit coupling $\lambda_\mathrm{SO}$ in this compound.

\section{Detailed CDW and SC Transition Curves and Peak Effect}
\begin{figure}
\includegraphics[width=0.9\linewidth]{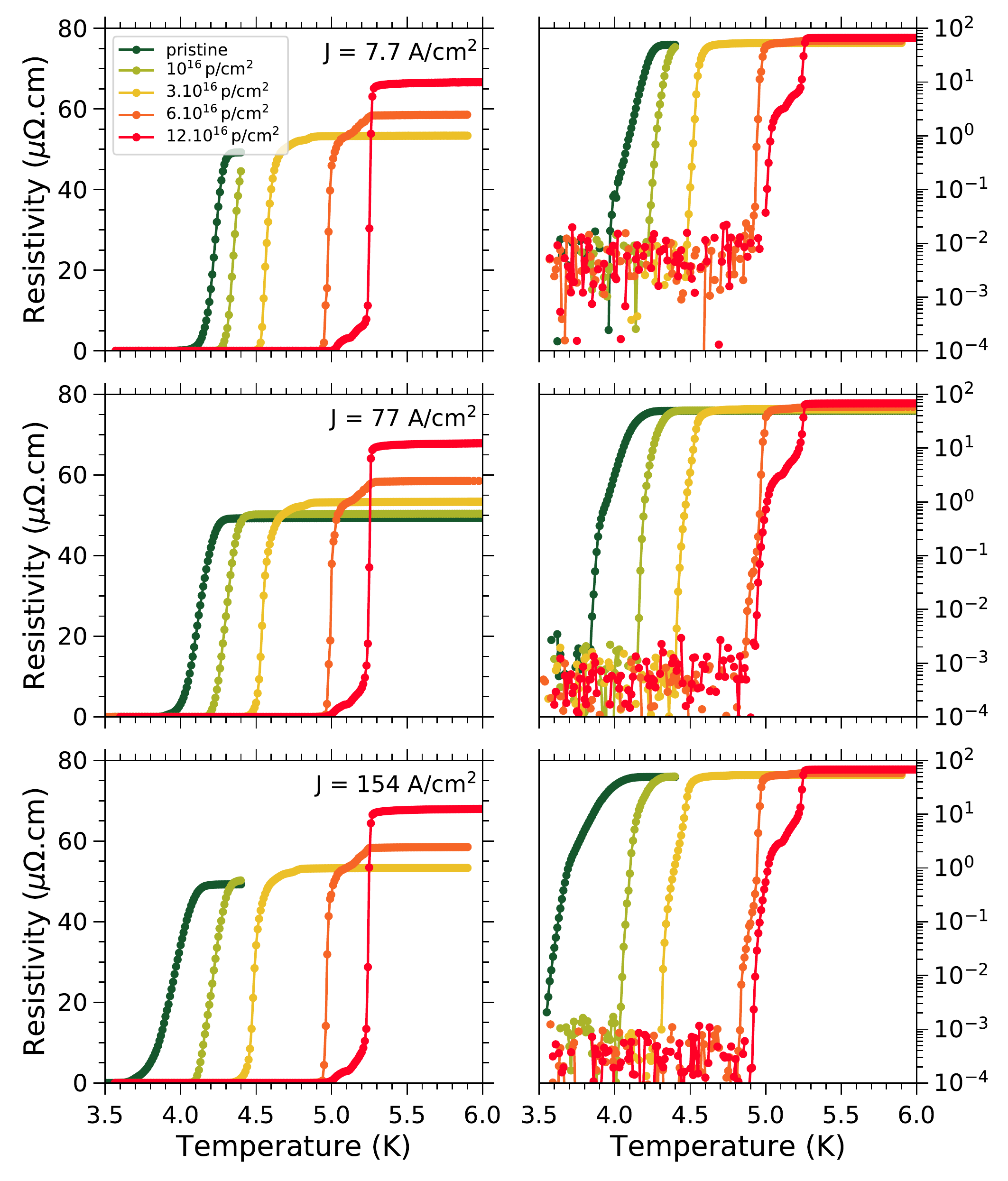}
\caption{\label{fig:RT_Tc_vsJ_log} \textbf{Detailed temperature dependence of the c-axis resistivity of Lu$_5$Ir$_4$Si$_{10}$ at the superconducting transition:} for increasing current densities (top to bottom) and in linear (left column) and semilog scale (right column).
}
\end{figure}

% Tc and current density
In Fig.~\ref{fig:RT_Tc_vsJ_log}, we show that the point where resistivity drops below the resolution of the instruments ($\approx10^{-3}\,\mu\Omega$.cm), also shifts to higher temperature with increasing irradiation dose, in the same manner as the midpoint of the transition.
We also note that in the pristine (non-irradiated) state, higher current densities shift the transition to lower temperature, whereas after irradiation the curves become almost identical at all current densities.
We found no effect of the current density on the CDW transition in the range of current density that we explored ($\leq 154$\,A/cm$^2$).

\begin{figure}
\includegraphics[width=0.9\linewidth]{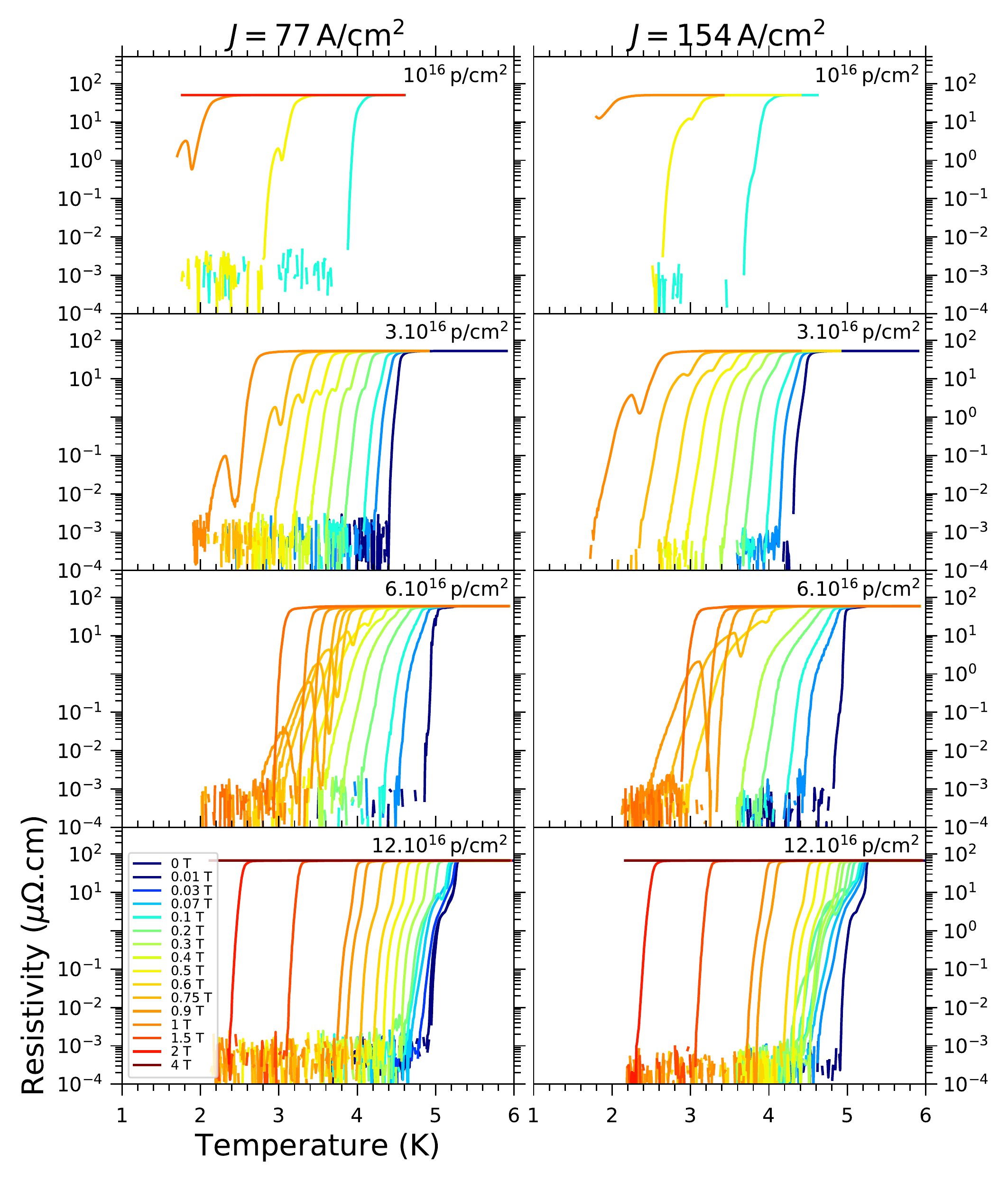}
\caption{\label{fig:RT_vsH_vsJ_log} \textbf{Detailed magnetic field dependence (H//ab) of the c-axis resistivity of Lu$_5$Ir$_4$Si$_{10}$ at the superconducting transition}: for increasing irradiation doses (top to bottom) and using a current density of 77 (left column) and 154 A/cm$^2$ (right column). A peak effect is visible in the superconducting transition.
}
\end{figure}

In Fig.~\ref{fig:RT_vsH_vsJ_log}, we show that we observe a clear peak effect in the middle of the superconducting transition, at both current densities and for all irradiation doses, which shows this peak effect is robust to disorder.
Such a peak effect\cite{FirstPeakEffect1961,Pippard1969PeakEffect,Larkin1979,Brandt2001PeakEffect,KOOPMANN2004} is usually caused by the softening of the vortex lattice near \Tc\ which enables it to better adapt to the distribution of defects, hence the drop in resistivity.
To our knowledge, such a peak effect had never been reported in \LIS.
We also find that the irreversibility field ($H_{irr}$) defined as the point where the resistivity drops below the resolution of our instruments ($1\,$n$\Omega$.cm) increases with irradiation dose, in line with $H_\mathrm{c,2}$.

\end{document}